\documentclass[prb,superscriptaddress,longbibliography,twocolumn]{revtex4-1}
\pdfoutput=1
\usepackage{geometry}
\usepackage{verbatim}
\usepackage{amsmath}
\usepackage{amssymb}
\usepackage{graphicx,picture,calc}
\usepackage{braket}
\usepackage{bm}
\usepackage{epstopdf}
\usepackage{xcolor}
\usepackage[multidot]{grffile}
\usepackage{overpic}
\usepackage{dsfont}
\newcommand{\be}{\begin{equation}}
\newcommand{\ee}{\end{equation}}

\newcommand{\bea}{\begin{eqnarray}}
\newcommand{\eea}{\end{eqnarray}}
\newcommand{\beal}{\begin{align}}
\newcommand{\eeal}{\end{align}}

\usepackage{newtxtext,newtxmath} % Times new Roman font.
\usepackage{hyperref} % Allows both internal and external hyperlinks

\usepackage{changes} % Conflicts with ulem.

\newcommand{\re}{\mathrm{e}} % roman e
\newcommand{\ri}{\mathrm{i}} % roman i
\newcommand{\rd}{\mathrm{d}} % roman d
\newcommand{\rpi}{\mathrm{\pi}} % roman pi
\newcommand{\rB}{\mathrm{B}} % roman B
\newcommand{\bE}{\mathbf{E}} % bold E
\newcommand{\bM}{\mathbf{M}} % bold M
\newcommand{\bQ}{\mathbf{Q}} % bold Q
\newcommand{\bR}{\mathbf{R}} % bold R
\newcommand{\ba}{\mathbf{a}} % bold a
\newcommand{\bb}{\mathbf{b}} % bold b
\newcommand{\bc}{\mathbf{c}} % bold c
\newcommand{\bj}{\mathbf{j}} % bold j
\newcommand{\bk}{\mathbf{k}} % bold k
\newcommand{\boldm}{\mathbf{m}} % bold m
\newcommand{\bq}{\mathbf{q}} % bold q
\newcommand{\br}{\mathbf{r}} % bold r
\newcommand{\bs}{\mathbf{s}} % bold s
\newcommand{\bv}{\mathbf{v}} % bold v
\newcommand{\bx}{\mathbf{x}} % bold x
\newcommand{\by}{\mathbf{y}} % bold y
\newcommand{\bz}{\mathbf{z}} % bold z
\newcommand{\bsigma}{\boldsymbol{\sigma}} % bold sigma
\newcommand{\vd}{\vec{d}} % vector tau
\newcommand{\vtau}{\vec{\tau}} % vector tau

\newcommand{\TRS}{\mathit{\Theta}} % time-reversal symmetry
\newcommand{\IS}{\mathit{\Pi}} % inversion symmetry

\renewcommand{\Im}{\text{Im\,}}

\begin{document}

\title{Orbital Edelstein effect from density-wave order}
\author{Geremia Massarelli}
\affiliation{Department of Physics, University of Toronto, Toronto, Ontario M5S 1A7, Canada}
\author{Bryce Wu}
\affiliation{Department of Physics, University of Toronto, Toronto, Ontario M5S 1A7, Canada}
\author{Arun Paramekanti}
\affiliation{Department of Physics, University of Toronto, Toronto, Ontario M5S 1A7, Canada}

\begin{abstract}
Coupling between charge and spin, and magnetoelectric effects more generally, have been an area of great interest for several years, with the sought-after ability to control magnetic degrees of freedom via charge currents serving as an impetus. The orbital Edelstein effect (OEE) is a kinetic magnetoelectric effect consisting of a bulk orbital magnetization induced by a charge current. It is the orbital analogue of the spin Edelstein effect in spin-orbit coupled materials, in which a charge current drives nonzero electron spin magnetization. The OEE has recently been investigated in the context of Weyl semimetals and Weyl metals. Motivated by these developments, we study a model of electrons without spin-orbit coupling which exhibits line nodes that get gapped out by via symmetry breaking due to an interaction-induced charge density wave order. This model is shown to exhibit a temperature dependent OEE, which appears due to symmetry reduction into a gyrotropic crystal class.
\end{abstract}

\maketitle

\section{Introduction}
The field of magnetoelectric effects has seen a revival in interest in the past decades on several fronts. The discovery and study of multiferroicity in correlated materials has uncovered unconventional mechanisms which can give rise to a large effective magnetoelectric coupling \cite{spaldin2005renaissance, eerenstein2006multiferroic, ramesh2007multiferroics, cheong2007multiferroics, spaldin2010multiferroics, lawes2011introduction, fuentescobas2015advances, chu2018review}. Similarly, the discovery of three-dimensional (3D) topological insulators has led to an exploration of novel magnetoelectric effects due to emergent axion electrodynamics in such topological phases \cite{hasan2010colloquium, qi2011topological, grushin2012finite, pesin2013topological, schmeltzer2013magnetoelectric, mal2013nonlinear, baasanjav2014magnetoelectric, morimoto2015topological, xiao2018realization, tokura2019magnetic}. A part of the reason for the wide interest in such magnetoelectric effects partly stems from the technological potential of controlling 
charge degrees of freedom via magnetic fields or,
conversely, tuning magnetic degrees of freedom via an applied electric field.

A prominent example of such a magnetoelectric effect is the nonequilibrium phenomenon of current-induced magnetization, which is also termed as \emph{kinetic magnetoelectric effect} (KME)\cite{levitov1985magnetoelectric, csahin2018pancharatnam}. The intrinsic-spin variant of this effect, wherein a charge current in a spin-orbit-coupled conductor gives rise to bulk spin polarization and, hence, a net magnetization, is referred to as the \emph{Edelstein effect} or the \emph{inverse spin-galvanic effect}\cite{sinova2015spin, manchon2015new} and has been under study for several decades\cite{ganichev2016spin}. A great deal of experimental work has focused on the Edelstein effect in 2D systems, notably in thin-film semiconductors\cite{kato2004current, silov2004current, kato2005electrical, sih2005spatial, stern2006current, yang2006spectral} and at metal surfaces\cite{zhang2014current}. However, experiments on 3D materials have been scant, although some recent studies have reported its observation in trigonal tellurium \cite{shalygin2012current, furukawa2017observation}.

In recent years, it has come to light that 3D systems can have an intrinsic orbital contribution to the KME, analogous to the spin part and arising as a consequence of the orbital magnetic moment of Bloch bands\cite{yoda2015current, zhong2016gyrotropic, rou2017kinetic, yoda2018orbital, csahin2018pancharatnam, souza2018gyrotropic, flicker2018chiral, chengwang2018mixed, shi2019symmetry}, notably in trigonal selenium and tellurium. 
%\emph{It is appropriate to emphasize that this magnetic moment is distinct from that of atomic orbitals and does not arise from atomic physics.} \textbf{[\emph{Double-check this statement.}]}
Whereas the ordinary Edelstein effect (hereafter referred to as the \emph{spin} Edelstein effect, SEE) relies on crystalline spin-orbit coupling (SOC) to give Bloch states a spin texture and, hence, is limited by the size of the SOC, the \emph{orbital Edelstein effect} (OEE), also referred to as the \emph{inverse gyrotropic magnetic effect}\cite{zhong2016gyrotropic}, is determined solely by the geometry of the crystal\cite{yoda2015current, yoda2018orbital}. 

Chiral crystals are a subset of those that can exhibit the KME. Previous studies have considered trigonal selenium and tellurium and viewed their chiral nature as descending from a charge-density-wave (CDW) instability of a hypothetical parent phase\cite{silva2018elemental}, and others have studied optical gyrotropy as a probe for symmetry breaking in the chiral CDW phase of $1T$-$\mathrm{TiSe_2}$ \cite{gradhand2015optical} and in stripe-ordered cuprates\cite{orenstein2013berry}. It has also been shown that Weyl nodes at the Fermi level can yield a large intrinsic contribution to the KME \cite{yoda2015current, yoda2018orbital}.

Our work builds on this theme, and explores the KME induced by symmetry breaking in a system with line nodes in the electronic band structure. The resulting phase is a non-chiral but gyrotropic crystal, and we study the concomitant temperature-dependent KME as a probe of the density-wave order. Below, we briefly review the OEE, before introducing our model Hamiltonian and presenting its theoretical study.

%Superconducting Edelstein effect\cite{fiebig2005revival}? Inverse Edelstein effect a.k.a.~spin-galvanic effect?

\section{Orbital Edelstein effect}

\begin{figure}[tb]
    \centering
    \includegraphics[width=0.75\linewidth]{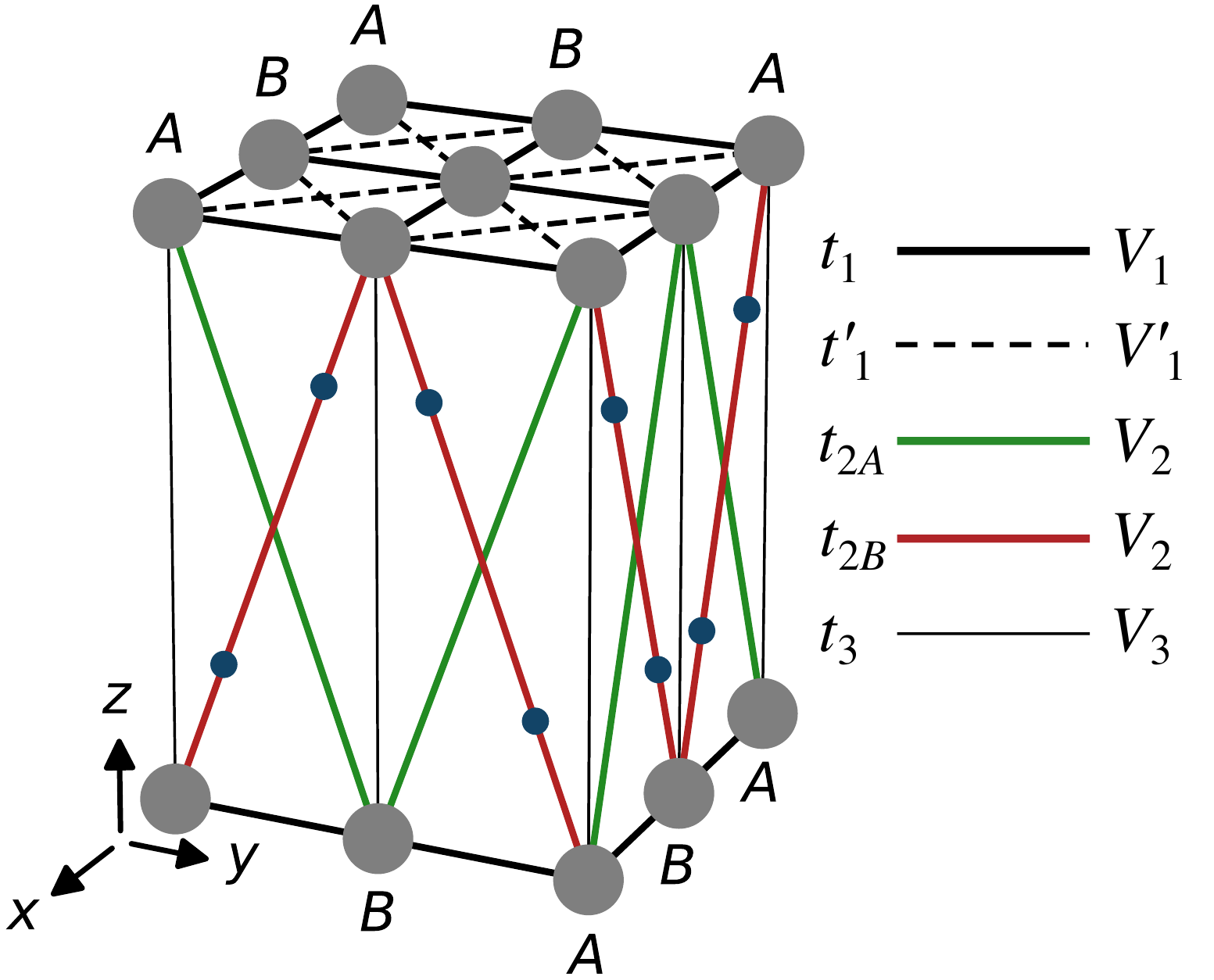}
    \caption{(Color online) Tetragonal crystal structure showing identical atoms (large gray spheres) on the two sublattices, with a legend for hopping amplitudes and repulsion strenghts. Primitive translations $\ba$, $\bb$, and $\bc$ are defined in the main text. The volume shown corresponds to two primitive lattice cells. The secondary atoms (small dark-blue spheres) are \emph{not} considered in the model---see text for a discussion.
    % In the broken-symmetry phase, the $A$ and $B$ atoms are distinguished by differing electron densities.
    }
    \label{fig:crystal_structure}
\end{figure}

Electrons in crystalline solids form Bloch bands with an intrinsic spin magnetic moment
\begin{equation}
\bs_{\bk n} = \frac{ge}{2m_\text{e}} \langle u_{\bk n} | \bs | u_{\bk n} \rangle,
\end{equation}
where $\bs = \hbar \bsigma/2$ is the spin operator for the electrons. The modern theory of magnetization in solids\cite{resta2010electrical, thonhauser2011theory, vanderbilt2018berry, resta2018electrical} has discovered that such Bloch bands also host an intrinsic orbital magnetic moment given by 
\begin{equation} \label{eq:Bloch_orbital_moment}
\boldm_{\bk n} = \frac{e}{2\hbar} \Im 
\langle \nabla_{\bk}u_{\bk n} | 
\times (H_{\bk} - \varepsilon_{\bk n}) 
| \nabla_{\bk}u_{\bk n} \rangle,
\end{equation}
where $H_{\bk}$ is the Bloch Hamiltonian, $n$ is the band index, and $H_{\bk} | u_{\bk n} \rangle = \varepsilon_{\bk n} | u_{\bk n} \rangle$. This orbital magnetization has been shown to arise from the self-rotation of wavepackets in the semiclassical theory of electron dynamics\cite{xiao2010berry, thonhauser2011theory}. 
%\textbf{[\emph{Explain quenching of OEE in 2D limit using wave packet picture?}]}  
Since $\boldm_{\bk n} \to -\boldm_{-\bk n}$ under time reversal and  $\boldm_{\bk n} \to \boldm_{-\bk n}$ under spatial inversion, it is clear that at least one of these symmetries must be broken in order for $\boldm_{\bk n}$ to not be identically zero.

From the viewpoint of semiclassical dynamics, given an electron distribution function $f_{\bk n}$, the instrinsic contribution to the net electronic magnetization is given by\cite{xiao2010berry, resta2010electrical, csahin2018pancharatnam}
\begin{equation}
\bM = \frac{1}{\mathcal{V}} \sum_{\bk n}  f_{\bk n} \left(\boldm_{\bk n}  + \bs_{\bk n}\right),
\end{equation}
where $\mathcal{V}$ is the crystal volume.
%There are additional contributions to $\bM$ arising as a result of disorder\cite{rou2017kinetic, csahin2018pancharatnam}; we ignore these so-called \emph{extrinsic} contributions in our treatment. \textbf{[\emph{Why?}]} 
In thermodynamic equilibrium for a time-reversal symmetric system, the Fermi-Dirac distribution, $f_{\bk n}^0 = f(\varepsilon_{\bk n}-\mu)$ forces zero a net magnetization $\bM=0$ because of cancellation between contributions from opposite crystal momenta\cite{yoda2018orbital}. However, an asymmetric distribution function, such as that arising from an applied electric field, can generally give rise to nonzero net bulk magnetization.

Explicitly, to lowest order in an applied uniform DC electric field, the distribution function is modified as\cite{ashcroft1976solid}
\begin{equation}
f_{\bk n} = f_{\bk n}^0 + e\tau (\bE\cdot\bv_{\bk n}) \left.\frac{\rd f}{\rd \xi}\right|_{\xi = \varepsilon_{\bk n}-\mu},
\end{equation}
where $\bv_{\bk n}$ is the electronic group velocity, $\tau$ is the impurity-scattering relaxation time in relaxation-time approximation, and $e>0$ is the elementary charge. Hence, the magnetization arises as a linear response to an applied electric field, 
\begin{equation}
M_\kappa = \alpha_{\kappa\lambda} E_\lambda,
\end{equation}
with the linear response tensor
\begin{subequations}\label{eq:alpha}
\begin{align}
\alpha_{\kappa\lambda}^{} &= \alpha_{\kappa\lambda}^\text{orb} + \alpha_{\kappa\lambda}^\text{spin} \\
&= \tau \frac{e}{\hbar}\sum_{\bk, n} 
\left.\frac{\rd f}{\rd \xi}\right|_{\xi = \varepsilon_{\bk n}-\mu}\!\!\!\!
(m_{\bk n, \kappa} + s_{\bk n, \kappa})~ v_{\bk n, \lambda},
\end{align}
\end{subequations}
where $\kappa$ and $\lambda$ are Cartesian indices.

The form of the tensor $\alpha$ is significantly constrained by crystal symmetry\cite{yoda2018orbital, ganichev2016spin, nye1957physical, landau2013electrodynamics}. $\alpha$ is an axial rank-two tensor since it relates a polar vector, $\bE$, to a axial vector, $\bM$. Crystal classes whose point-group symmetries allow for nonzero axial rank-two response tensors are known as \emph{gyrotropic}. The reason for this name is that the tensor governing natural optical activity, or \emph{gyrotropy}, transforms in the same way as $\alpha$; thus KME and optical gyrotropy go hand in hand. 

We note that the same symmetry constraints govern the appearance of nonzero spin and orbital contributions to $\alpha$, so both are expected to arise together, and there is no clear route to disentangling them in a 3D system\cite{yoda2018orbital, furukawa2017observation}. Indeed, the authors of Ref.~\onlinecite{furukawa2017observation} conclude by speculating that the current-induced magnetization they observe in trigonal tellurium may be due not only to the well-known SEE, but also to the OEE.

\section{Model} \label{sec:model}

\begin{figure}[tb]
\includegraphics[width=0.55\linewidth]{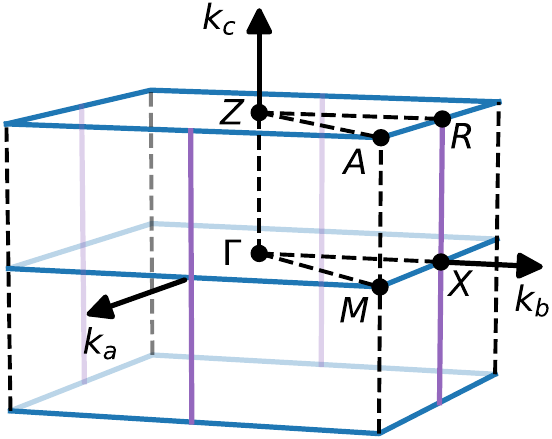}
\caption{(Color online) Location of the line nodes (solid lines, shown in color) present in the symmetric phase within the first Brillouin zone for the crystal under study. 
%when inversion symmetry remains unbroken, 
High-symmetry points are labeled.}
\label{fig:line_nodes}
\end{figure}

As an example of the OEE brought about by symmetry breaking, we consider a tight-binding toy model of spinless fermions moving in a tetragonal crystal as shown in Fig.~\ref{fig:crystal_structure}, consisting of identical atoms are arranged in layered square lattices. We assume a single isotropic orbital, and ignore the electron spin below; there are many cases where this is a useful starting point. In crystals with density-wave order driven by nearest-neighbor repulsion, such as we will consider below, both spin components behave in the same manner. Including spin then only leads to an extra factor-of-two in certain equations below. The other case where spin may be ignored is in spin-polarized systems which might be a useful description of states in a large energy interval around the Fermi  energy in strong ferromagnets (i.e., in half-metals).

We define the nearest-neighbor (NN) lattice constant $a_0$, and lattice constant along the stacking-$\hat{\bz}$ axis to be $c$. In the $x$-$y$ planes, we include NN hopping $t_1$ and next-nearest-neighbor (NNN) hopping $t_1'$. In the $y$-$z$ and $z$-$x$ planes, we include NN hopping $t_3$ and the peculiar NNN hoppings depicted in Fig.~\ref{fig:crystal_structure}, with $t_{2A}\neq t_{2B}$. This last choice differentiates staggered $A$ and $B$ sublattices, and endows the crystal with primitive translations $\ba = a_0(\hat{\bx} + \hat{\by})$, $\bb = a_0(\hat{\bx} - \hat{\by})$, and $\bc = c\hat{\bz}$. The crystal space group is $\mathit{P4/nbm}$, and its associated crystal class is $\mathit{4/mmm}$. Importantly, this crystal has centers of inversion at the middle point of every NN bond.

The inequivalence of the hopping amplitudes $t_{2A}$ and $t_{2B}$ for $\mathit{P4/nbm}$ symmetry may be rationalized as depicted in Fig.~\ref{fig:crystal_structure}: if a secondary set of atoms is present only along the red bonds---which would be consistent with the space-group symmetry of the model---and their energy levels are far from the Fermi energy, their dynamics could be integrated out, with the end result of renormalizing the hopping amplitude between the primary atoms (shown in gray), effectively leading to $t_{2A}\neq t_{2B}$. The same reasoning holds if there are additional secondary atoms on the green bonds \emph{provided they are of a species different from those on the red bonds}.

We use the following values for the hopping parameters: $t_1'=0.7\,t_1$, $t_{2A}=0.1\,t_1$, $t_{2B}=0.4\,t_1$, and $t_{3}=0.5\,t_1$.

As is true for any crystal class with inversion symmetry, $\mathit{4/mmm}$ is non-gyrotropic\cite{nye1957physical, landau2013electrodynamics}. However, if the symmetry of the crystal were to be reduced, for instance by the onset of CDW order, the inversion symmetry could be broken and, indeed, the ordered structure could fall into a gyrotropic class. Below, we first study the band structure of this model in the absence of interactions, before turning to the impact of nearest-neighbor repulsion.

\begin{figure}[tb]
\makebox[0.1\linewidth][c]{\includegraphics[width=0.62\linewidth]{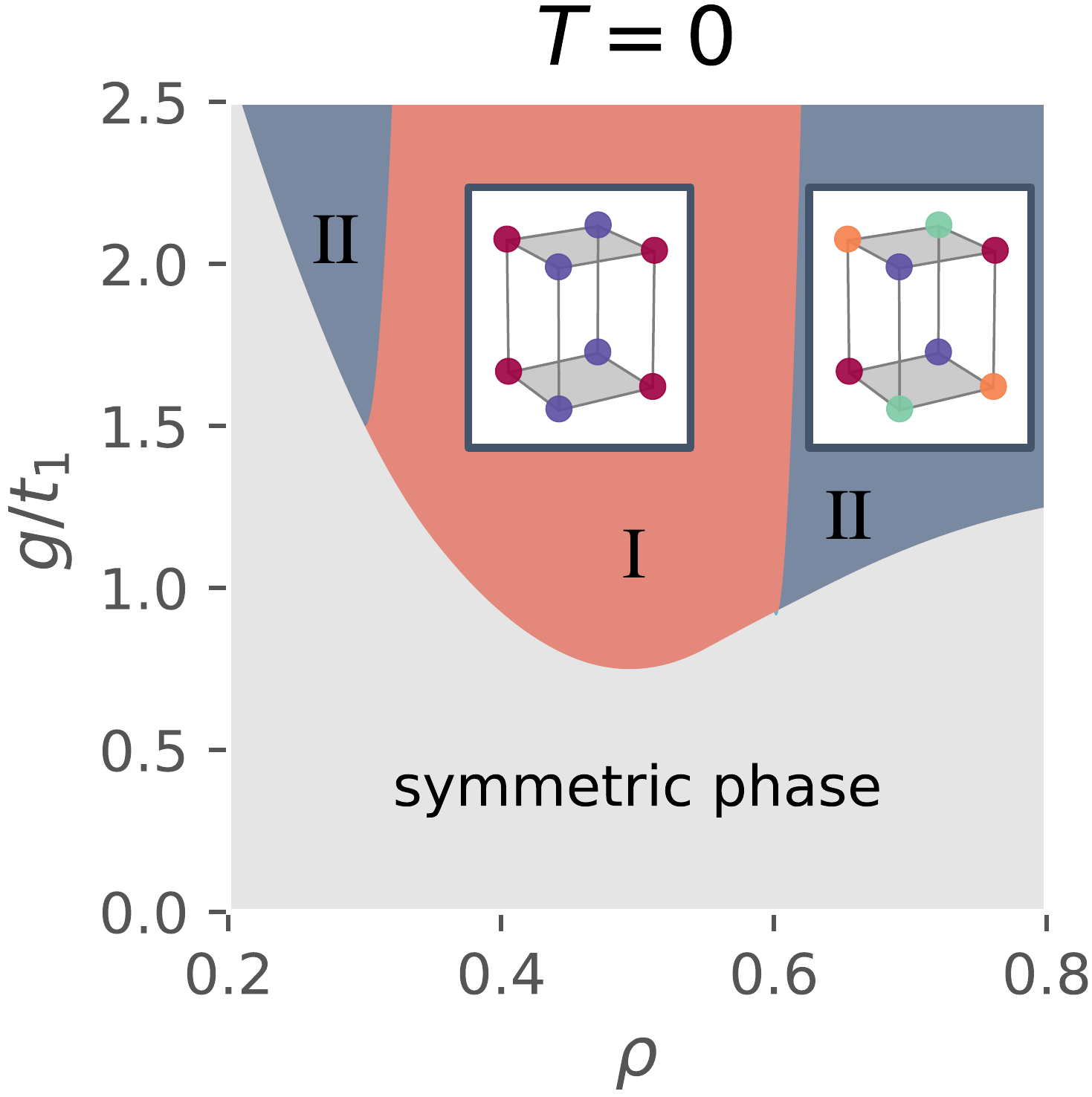}}
\caption{(Color online) CDW phases arising in the MF study of the Hamiltonian $H = K + V$ at $T=0$ as a function of the electronic filling $\rho$ and
$g$ (which parametrizes the repulsion strength). The nature of phases I and II is described in the text.}
\label{fig:phasediagram1}
\end{figure}

\subsection{Non-interacting band structure}

The non-interacting Hamiltonian is $K= \sum_{i,j}  - t_{ij}^{\alpha\beta} {c_i^\alpha}^\dagger c_j^{\,\beta}$, where ${c_i^\alpha}^\dagger$ creates an electron on sublattice $\alpha$ of unit cell $i$ (that is, on the atom at position $\br_i^\alpha$). In momentum space
\begin{equation} \label{eq:TB}
K = \sum_{\bk}
{\psi_{\bk}}^\dagger
\left(d_{\bk}^0 + \vd_{\bk}^{\vphantom{0}} \cdot \vtau \right)
\psi_{\bk},
\end{equation}
$c_{\bk}^\alpha = N^{-1/2} \sum_{i} \re^{-\ri \bk \cdot \br_i^{\alpha}} c_i^\alpha$, $N$ is the number of unit cells, $\vtau$ is the vector of Pauli matrices (acting in sublattice space), and ${\psi_{\bk}}^\dagger = \Big( {c_{\bk}^A}^\dagger \ \ {c_{\bk}^B}^\dagger \Big)$. Because of time-reversal symmetry, the hopping amplitudes $t_{ij}^{\alpha\beta}$ are necessarily real valued.
We will measure momenta in units of inverse lattice spacing, and henceforth set $a=c=1$. With this, we arrive at
\begin{subequations}
\begin{align}
d_{\bk}^0 &= -2 t_3 \cos(k_c) - 2 t_1' \Big( \cos(k_a) + \cos(k_b) \Big)\\
d_{\bk}^1 &= -4 \cos\left(\frac{k_a}{2}\right) \cos\left(\frac{k_b}{2}\right) \Big(t_1 +
 (t_{2A} + t_{2B}) \cos(k_c)
 \Big)\\
d_{\bk}^2 &= 4 (t_{2A} - t_{2B}) \cos\left(\frac{k_a}{2}\right) \cos\left(\frac{k_b}{2}\right) \cos(k_c) \\
d_{\bk}^3 &= 0.
\end{align}
\end{subequations}

The inversion symmetry $\IS$ mentioned above combined with the Hamiltonian's time-reversal symmetry $\TRS$ constrain $d^3_{\bk} \equiv 0$ identically; for this reason, band touchings for this crystal will generically arise as line nodes, as seen in the non-interacting band structure (top panel of Fig.~\ref{fig:MF_BS}) Figure~\ref{fig:line_nodes} shows the location of the line nodes in the Brillouin zone. 

Spontaneous symmetry breaking (SSB), however, could change the crystal class to a less symmetric one. A simple scenario is a CDW phase in which the densities on the $A$ and $B$ atoms are inequal, in which case the space group becomes $\mathit{P\bar{4}2m}$, whose crystal class---$\mathit{\bar{4}2m}$---is gyrotropic.

\begin{figure}[tb]
\makebox[0.1\linewidth][c]{\includegraphics[width=0.62\linewidth]{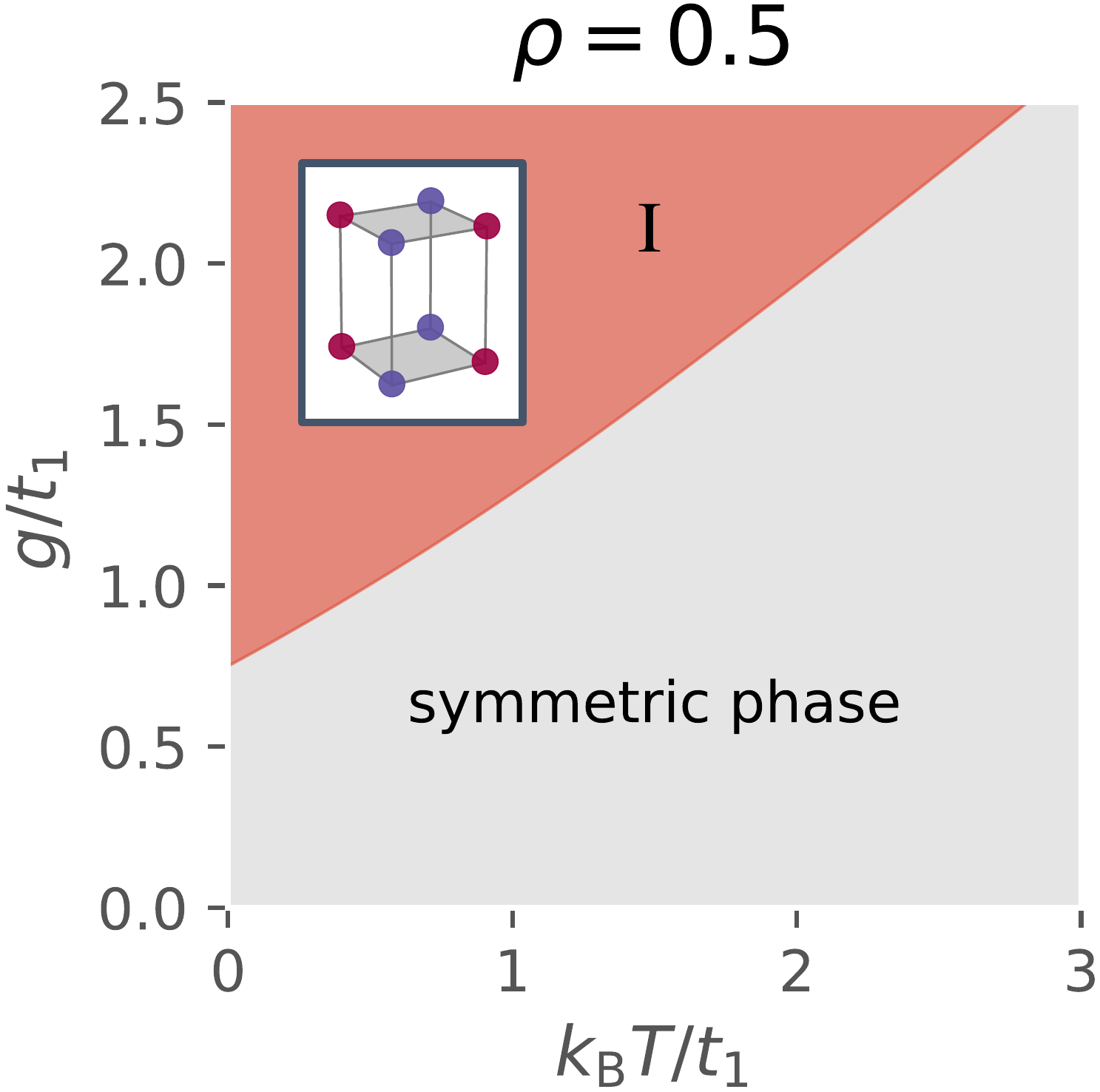}}
\caption{(Color online) CDW phases arising in the MF study of the Hamiltonian $H = K + V$ at half filling ($\rho = 0.5$) as a function of temperature and
$g$ (which parametrizes the repulsion strength). The nature of phase I is described in the text.}
\label{fig:phasediagram2}
\end{figure}

\subsection{Repulsive interactions} \label{subsec:repulsive_interactions}
Next, we include repulsive interactions between NNs and next-nearest neighbors (NNNs); that is,
\begin{equation} \label{eq:interaction}
V = \frac{1}{2} \sum_{(i,\alpha)\neq (j,\beta)} V_{ij}^{\alpha\beta} n_i^\alpha n_j^\beta,
\end{equation}
where $n_i^\alpha = {c_i^{\alpha}}^\dagger c_i^{\alpha}$ is the number operator for the atom $(i,\alpha)$ and we take $V_{ij}^{\alpha\beta} = V_{ji}^{\beta\alpha}$. We take nonzero repulsions $V_{ij}^{\alpha\beta}$ on the same bonds as the hopping amplitudes and use a similar naming scheme, as shown in Fig.~\ref{fig:crystal_structure}. Note that, unlike for the hopping amplitudes $t_{2A} \neq t_{2B}$, we take the repulsion strengths to be $V_2$ on both the red and green bonds --- we make this choice since it serves as a representative slice through the full parameter space and serves to illustrate some of the key ideas. Hence, the full Hamiltonian is given by $H = K + V$.

In the absence of spin order, a Hubbard term $U\sum_{i}n_{i\uparrow}n_{i\downarrow}$ would modify the MF Hamiltonian by a mere uniform offset in the chemical potential $\mu$. Hence, as we will restrict ourselves to ans\"atze without spin order, we do not include the Hubbard interaction.

We performed a MF calculation of the CDW order in the above system, allowing for a finite set of commensurate wavevectors. We identified the ordering wavevectors favoured by the interaction by considering a simple model of classical charges resting at each atomic site of the crystal of Fig.~\ref{fig:crystal_structure}---see Appendix~\ref{app:charge_orders}. Based on this, we included in our ansatz the four ordering wavevectors 
\begin{align*}
\bQ_0 &= 0 &
\bQ_1 &= \rpi\hat{\ba} + \rpi\hat{\bb} \\
\bQ_2 &= \rpi\hat{\bc} &
\bQ_3 &= \rpi\hat{\ba} + \rpi\hat{\bb} + \rpi\hat{\bc}.
\end{align*}
The order parameters for the mean-field theory are the Fourier amplitudes $\rho_{\bQ}^\alpha$ for the $\bQ$ listed above, where $\alpha \in \{A,B\}$. However, we find it convenient to express the amplitudes on the $A$ and $B$ sublattices in terms of a symmetric and an antisymmetric part, respectively defined as
\begin{subequations}
\begin{align}
\rho_{\bQ}^\text{s} &:= \frac{\rho_{\bQ}^A + \rho_{\bQ}^B}{2}, \\
\rho_{\bQ}^\text{a} &:= \frac{\rho_{\bQ}^A - \rho_{\bQ}^B}{2}.
\end{align}
\end{subequations}
Since $\rho_{\bQ_0}^\text{s} = \rho$, the electron filling, we are left with 7 independent MFs.

\section{Results} \label{sec:results}

\subsection{Mean-field theory of charge-density-wave order}

We focus on a cut of parameter space parametrized by $g$ such that
\begin{align}
V_1 &= g, & 
V_1' &= \frac{g}{2}, &
V_2 &= \frac{g}{2}, &
V_3 &= \frac{g}{2}.
\end{align}
According to the model of electrostatic charges, at these relative repulsion values (marked by a white star in Fig.~\ref{fig:classical_phase_diagram}), the interaction most favours an ordering with $\rho_{\bQ_0}^\text{a} \neq 0$; however, the region of parameter space with $\rho_{\bQ_3}^\alpha \neq 0$ most favoured is nearby (see Appendix~\ref{app:charge_orders}). 

The MF calculation, for which the zero-temperature phase diagram is shown in Fig.~\ref{fig:phasediagram1}, does indeed identify a swath of pure $\rho_{\bQ_0}^\text{a} \neq 0$ order for $g$ sufficiently large and for approximately $0.3 < \rho < 0.6$---call this order \emph{phase I}. In addition to this phase, we discover mixed phases at high and low filling in which both $\rho_{\bQ_0}^\text{a} \neq 0$ and $\rho_{\bQ_3}^\alpha \neq 0$, where $\alpha$ is either $A$ or $B$---call the order in these regions \emph{phase II}. 

For $g/t_1 \gg 1$, we observe that the phase diagram is approximately symmetric under $\rho \rightarrow 0.5 - \rho$; this is expected given the particle-hole symmetry of $V$, which becomes an approximate symmetry of $H$ when $K$ is nonzero but small compared with $V$.

The phase diagram at $\rho=0.5$ as a function of $T$ and $g$ is shown in Fig.~\ref{fig:phasediagram2} and reveals that raising the system temperature progressively supresses phase-I order.

\begin{figure}[tb]
\makebox[0.1\linewidth][c]{\includegraphics[width=1.0\linewidth]{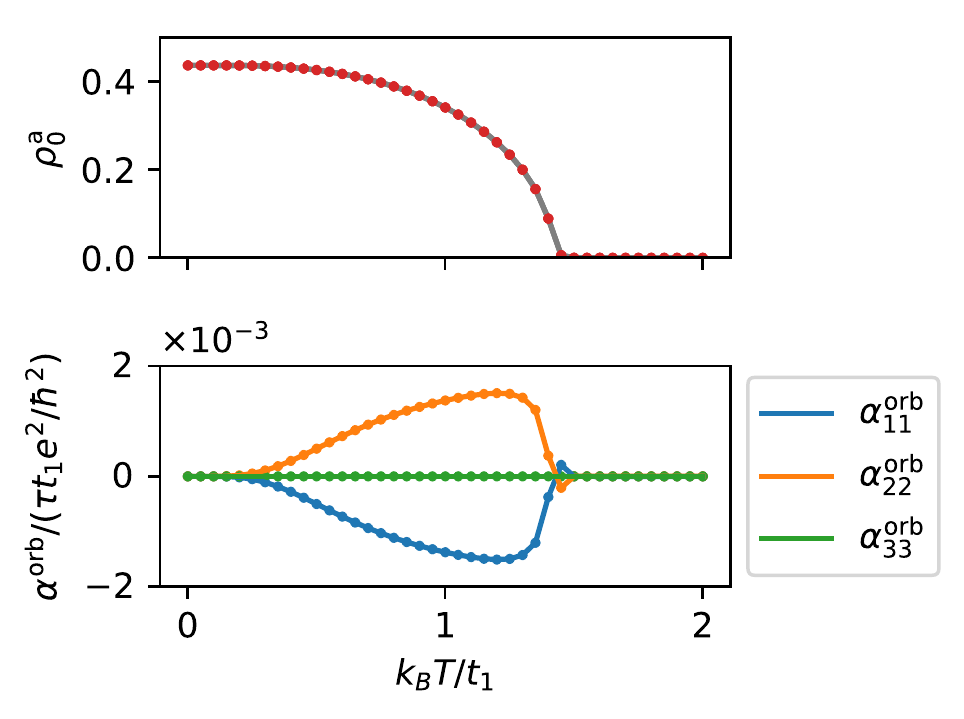}}
\caption{(Color online) Evolution of the magnetoelectric response through a phase transition with $g = 1.5 t_1$ and $\rho = 0.5$. \textsc{Top panel} Order parameter $\Big(\rho_{\bQ_0}^\text{a}\Big)$ as a function of temperature.\quad \textsc{Bottom panel} Components of the OEE response tensor $\alpha^\text{orb}$ as a function of temperature. Components that are not shown are identically zero.}
\label{fig:magnetoelectric_response}
\end{figure}

\subsection{Magnetoelectric response in phase I}
We studied the temperature-dependent SSB-induced OEE in broken-symmetry phase I, whose sole order parameter is $\rho_{\bQ_0}^\text{a}$. Figure~\ref{fig:magnetoelectric_response} shows the evolution of $\rho_{\bQ_0}^\text{a}$ and the components of the response tensor $\alpha^\text{orb}$ at half filling ($\rho = 0.5$) and fixed interaction strength ($g = 1.5\,t_1$). The phase transition, which is seen to be continuous, takes place at $k_\text{B}T_\text{c} \approx 1.4\,t_1$. 

\begin{figure}[tb]
    \centering
    \makebox[0.1\linewidth][c]{\includegraphics[width=1.1\linewidth]{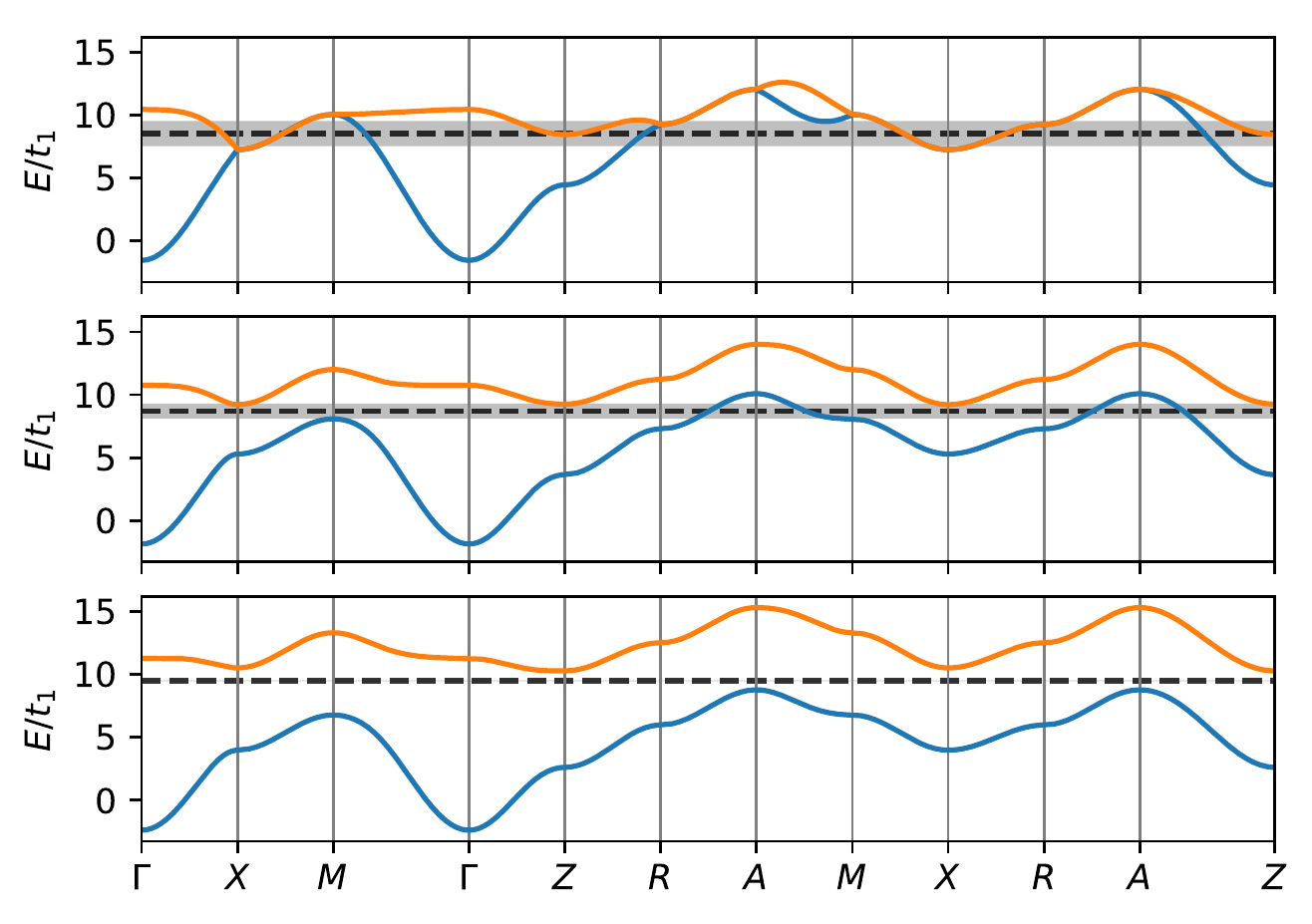}}
    \caption{(Color online) Mean-field band structure at half filling (with parameters as specified in the main text) at three different temperatures: $k_{\rB}T/t_1 = 2.00$ (top), $1.20$ (middle), and $0.05$ (bottom). The black dotted lines show the location of the chemical potential $\mu$ at each temperature, and the shaded areas centered at $\mu$ show the width $k_{\rB}T$. 
    %In the first case is an insulating phase, giving $\alpha=0$. The last case is in the higher-symmetry, non gyrotropic phase---hence the presence of band touchings---and thus $\alpha=0$. The intermediate case gives nonzero magnetoelectric response.
    }
    \label{fig:MF_BS}
\end{figure}

As alluded to previously, the point-group symmetry in phase I is the $\mathit{\bar{4}2m}$, and a symmetry analysis reveals that symmetry constrains the response tensor $\alpha$ to the form\cite{nye1957physical}
\begin{equation} \label{eq:alpha_symmetry}
\alpha = \begin{pmatrix}
\alpha_{aa} & & \\
& -\alpha_{aa} & \\
&&0
\end{pmatrix}.
\end{equation}
As seen in the lower panel of Fig.~\ref{fig:magnetoelectric_response}, the calculated $\alpha^\text{orb}$ is indeed of this form. The nonmonotonicity of $\alpha^\text{orb}$ is understood via the MF band structure, seen in Fig.~\ref{fig:MF_BS}. The top panel shows the high-temperature ($k_\text{B}T = 2.00\,t_1$) band structure. Although the system is in a metallic phase with many states within $k_\text{B}T$ (gray shading) from the chemical potential $\mu$ (dotted line), in the high-symmetry phase, $\boldm_{\bk n}=0$ identically, so $\alpha^\text{orb}=0$. The middle panel shows the band structure at an intermediate temperature $k_\text{B}T = 1.20\,t_1$ (near the peak in the $\alpha^\text{orb}$); the phase-I charge order has broken inversion symmetry and split the bands throughout the Brillouin zone, and the reduced symmetry of phase allows gyrotropic response, so $\alpha^\text{orb}\neq0$. As the temperature is further lowered, the size of the band splitting increases and a full gap develops, with $\mu$ within this gap (lower panel of Fig.~\ref{fig:magnetoelectric_response}). Hence, it is clear that in the low-temperature limit, as the sum in Eq.~\ref{eq:alpha} approaches a Fermi-surface integral, $\alpha^\text{orb}$ should again vanish.

%\subsection{Line-node semimetal phase}
\subsection{Role of line nodes}

\begin{figure*}
\centering
    \makebox[0.1\linewidth][c]{\includegraphics[width=1.1\linewidth]{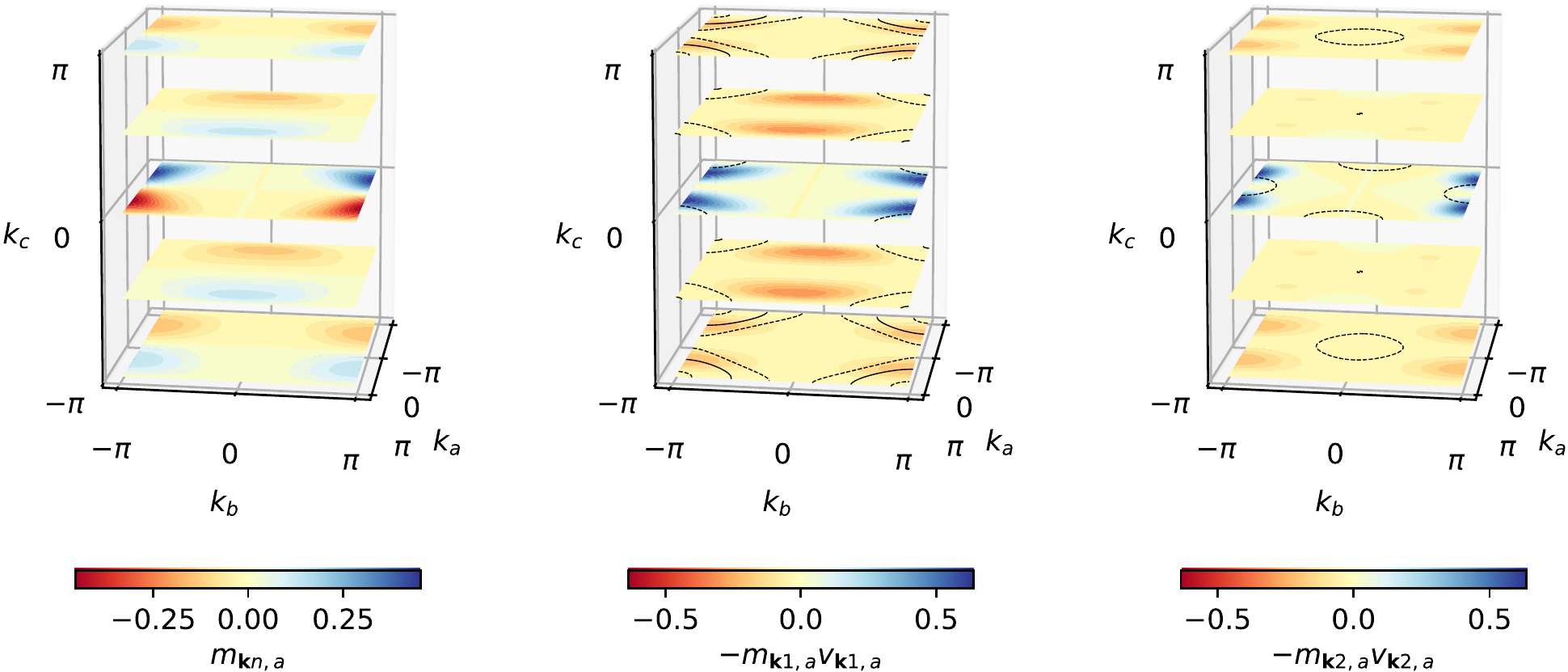}}
    \caption{(Color online) Magnetic moment $m_{\bk n, a}$ and product $-m_{\bk n, a}v_{\bk n, a}$ for bands $n=1$ and $2$ at $k_\text{B}T=1.20\,t_1$. As expected, $m_{\bk n, a}$ is odd under $k_a \to - k_a$. The solid lines in middle and right panels depict $\varepsilon_{\bk n}=\mu$ surfaces, while the dashed lines show $\varepsilon_{\bk n}=\mu\pm k_\text{B}T$ surfaces. Other components are either related by symmetry or vanish.}
    \label{fig:integrand}
\end{figure*}

In this subsection, we inspect the contribution of the gapped-out line nodes to the magnetoelectric response tensor $\alpha^\text{orb}$ in phase I. As depicted in Fig.~\ref{fig:line_nodes}, in the high-symmetry phase, three independent line nodes exist in the first Brillouin zone: 1.~the $X$-$R$ segment, 2.~the $X$-$M$ segment, and 3.~the $R$-$A$ segment---all other line nodes are related by symmetry.

In phase I, the charge order breaks the inversion symmetry $\IS$; accordingly, in the MF Hamiltonian for phase I, $d_{\bk}^3$ is no longer constrained to be zero, so the band touchings disappear. Rather, $d_{\bk}^3$ is set by the phase-I order parameter $\rho_{\bQ_0}^\text{a}$. As a shorthand, define
\begin{equation}
\nu := \tilde{\rho}_{\bQ_0}^\text{a},
\end{equation}
where $\tilde{\rho}_{\bQ_0}^\text{a}$, defined in Appendix~\ref{app:MFT}, is proportional to the phase-I order parameter $\rho_{\bQ_0}^\text{a}$.

We expand about the (gapped) line nodes, denoting small deviations in momenum by $\tilde{\bk}$. In the neighborhood of the (gapped) nodes, we have
\begin{enumerate}
\item $k_a = \rpi + \tilde{k}_a$,\quad $k_b = \tilde{k}_b$,\quad $-\rpi < k_c \leq \rpi$
\begin{subequations}
\begin{align}
d_{\bk}^0 &= -2t_3 \cos(k_c) \\
d_{\bk}^1 &= 2 \big( t_1 + (t_{2A} + t_{2B})\cos(k_c) \big) \tilde{k}_a \\
d_{\bk}^2 &= 2(t_{2A} - t_{2B})\sin(k_c) \tilde{k}_b \\
d_{\bk}^3 &= \nu
\end{align}
\end{subequations}

%\item (Old)
%$k_a = \rpi + \tilde{k}_a$,\quad $-\rpi < k_b \leq \rpi$,\quad $k_c = \tilde{k}_c$
%\begin{subequations}
%\begin{align}
%d_{\bk}^0 &= -2t_3 + 2t_1' \big(1 - \cos(k_b)\big) \\
%d_{\bk}^1 &= 2(t_1 + t_{2A} + t_{2B}) \cos\left(\frac{k_b}{2}\right) \tilde{k}_a \\
%d_{\bk}^2 &= 4(t_{2A} - t_{2B})\sin\left(\frac{k_b}{2}\right) \tilde{k}_c \\
%d_{\bk}^3 &= \nu
%\end{align}
%\end{subequations}

\item $-\rpi<k_a<\rpi, \quad k_b = \rpi + \tilde{k}_b, \quad k_c=\tilde{k}_c$
\begin{subequations}
\begin{align}
d_{\bk}^0 &= -2t_3 + 2t_1' \big(1 - \cos(k_a)\big) \\
d_{\bk}^1 &= 2(t_1 + t_{2A} + t_{2B}) \cos\left(\frac{k_a}{2}\right) \tilde{k}_b \\
d_{\bk}^2 &= 4(t_{2A} - t_{2B})\sin\left(\frac{k_a}{2}\right) \tilde{k}_c \\
d_{\bk}^3 &= \nu
\end{align}
\end{subequations}

%\item (Old)
%$k_a = \rpi + \tilde{k}_a$,\quad $-\rpi < k_b \leq \rpi$,\quad $k_c = \rpi + \tilde{k}_c$
%\begin{subequations}
%\begin{align}
%d_{\bk}^0 &= 2t_3 + 2t_1' \big(1 - \cos(k_b)\big) \\
%d_{\bk}^1 &= 2(t_1 - t_{2A} - t_{2B}) \cos\left(\frac{k_b}{2}\right) \tilde{k}_a \\
%d_{\bk}^2 &= -4(t_{2A} - t_{2B})\sin\left(\frac{k_b}{2}\right) \tilde{k}_c \\
%d_{\bk}^3 &= \nu
%\end{align}
%\end{subequations}

\item $-\rpi<k_a<\rpi, \quad k_b = \rpi + \tilde{k}_b, \quad k_c = \rpi + \tilde{k}_c$
\begin{subequations}
\begin{align}
d_{\bk}^0 &= 2t_3 + 2t_1' \big(1 - \cos(k_a)\big) \\
d_{\bk}^1 &= 2(t_1 - t_{2A} - t_{2B}) \cos\left(\frac{k_a}{2}\right) \tilde{k}_b \\
d_{\bk}^2 &= -4(t_{2A} - t_{2B})\sin\left(\frac{k_a}{2}\right) \tilde{k}_c \\
d_{\bk}^3 &= \nu.
\end{align}
\end{subequations}
\end{enumerate}

We can compute the orbital magnetic moment about such a (gapped) line node: for a two-band model with Bloch Hamiltonian $H_{\bk} = d_{\bk}^0 \tau^0 + \vd_{\bk}\cdot\vtau$, $\boldm_{\bk n}$ can be written in the closed form\cite{zhong2016gyrotropic, yoda2018orbital}
\begin{equation}
m_{\bk n, \kappa} = - \frac{e}{2\hbar} 
\varepsilon_{\kappa\lambda\zeta} \frac{\vd_{\bk}}{\big|\vd_{\bk}\big|^2} \cdot \left(\frac{\partial\vd_{\bk}}{\partial k_\lambda} \times \frac{\partial\vd_{\bk}}{\partial k_\zeta}\right),
\end{equation}
where $\kappa$, $\lambda$, and $\zeta$ are Cartesian indices. 

In the neighborhood of a (gapped) line node in the $\hat{\ba}$ direction,
\begin{align}\label{eq:line_node_ham}
d^0_{\bk} &= f(k_a), &
\vd_{\bk} &= \left( g_1(k_a)\tilde{k}_b,\ g_2(k_a)\tilde{k}_c,\ h(k_a) \right),
\end{align}
where $f$, $g_1$, $g_2$, and $h$ are functions of $k_a$, and, hence, the components of $\boldm_{\bk n}$ are given by
\begin{subequations}\label{eq:line_node_mag}
\begin{align}
m_{\bk n, a} &= \frac{e}{2\hbar} \frac{1}{\big|\vd_{\bk}\big|^2} g_1 g_2 h \\
m_{\bk n, b} &= \frac{e}{2\hbar} \frac{\tilde{k}_b}{\big|\vd_{\bk}\big|^2} g_2 (g_1 h' - g_1' h) \\
m_{\bk n, c} &= \frac{e}{2\hbar} \frac{\tilde{k}_c}{\big|\vd_{\bk}\big|^2} g_1 (g_2 h' - g_2' h),
\end{align}
\end{subequations}
where primes denote differentiation of the single-variable functions. To describe a line node in another direction, the indices in Eqs.~\ref{eq:line_node_ham} and \ref{eq:line_node_mag} must be interchanged appropriately.

Figure~\ref{fig:integrand} shows that the main contributions to $\alpha^\text{orb}$ originate from the the line nodes. The figure, for which we take $k_\text{B}T = 1.20\,t_1$ as in the middle panel of Fig.~\ref{fig:MF_BS}, shows the orbital magnetic moment $m_{\bk n, a}$ (left panel) as well as the product $-m_{\bk n, a} v_{\bk n, a}$ for the bottom ($n=1$, middle panel) and top ($n=2$, top panel) bands, which enter into the calculation of $\alpha^\text{orb}_{aa}$.
%For the purpose of the figure, we focus on the component $\alpha^\text{orb}_{aa}$
[The component $\alpha^\text{orb}_{bb}$ is related by symmetry (Eq.~\ref{eq:alpha_symmetry}), while all other components vanish.] These quantities are shown along slices $k_c=-\rpi$, $-\rpi/2$, $0$, $\rpi/2$, and $\rpi$. 

The form of $m_{\bk n, a}$ in momentum space reflects all the symmetries of the point group $\mathit{\bar{4}2m}$ of phase I. The large values of $m_{\bk n, a}$ are attributable to the presence of line nodes and are as expected fom Eqs.~\ref{eq:line_node_mag}: for the slices $k_c=\pm\rpi/2$, we see the signature of the line node transverse to $\hat{\ba}$, for which $m_{\bk n, a}\rightarrow 0$ in proximity to the node, and for the slices $k_c=0$ and $k_c=\pm\rpi$, we also recognize contributions from the line nodes longitudinal to $\hat{\ba}$, which are largest at the locus of the line nodes. The product $-m_{\bk n, a} v_{\bk n, a}$ still retains the general form of $m_{\bk n, a}$, though with a modified symmetry that allows a nonzero integral over the Fermi surface. 

The difference in the magnitude of the longitudinal contributions at $k_c=0$ and $k_c=\pm\rpi$ can be retraced to the different $g_1(k_a)$ for the line nodes in those planes: while $g_1(k_a)=2(t_1+t_{2A}+t_{2B})\cos(k_a/2)$ for the former, $g_1(k_a)=2(t_1-t_{2A}-t_{2B})\cos(k_a/2)$ for the latter, and $|t_1+t_{2A}+t_{2B}|>|t_1-t_{2A}-t_{2B}|$ with our parameter values.

As shown by Eq.~\ref{eq:alpha}, up to constant factors, the integrand for $\alpha^\text{orb}_{aa}$ is $-m_{\bk n, a} v_{\bk n, a}$ multiplied by $-\rd f / \rd \xi|_{\xi=\varepsilon_{\bk n}-\mu}$, a weighting factor concentrated within $k_\text{B}T$ around the chemical potential $\mu$. Hence, the equal-energy surfaces at $\varepsilon_{\bk n}=\mu$ (solid black lines) and $\varepsilon_{\bk n}=\mu\pm k_\text{B}T$ (dashed black lines) reveal the main contributions to the integral. At this temperature, for our choice of parameters, the largest contributions to $\alpha^\text{orb}$ are from the (gapped) line nodes at $k_c=\pm\rpi$.

\section{Discussion and conclusion}

It is sometimes overlooked that gyrotropy can arise without chiral or time-reversal symmetry breaking\cite{orenstein2013berry, ganichev2016spin}.
%; indeed, there are six non-chiral gyrotropic crystal classes: $\mathit{m}$, $\mathit{2mm}$, $\mathit{3m}$, $\mathit{\bar{4}}$, $\mathit{\bar{4}2m}$, $\mathit{4mm}$, and $\mathit{6mm}$.
Since the point group $\mathit{\bar{4}2m}$ of the phase I order contains mirror symmetries (in our coordinates, one perpendicular to $\hat{\bx}$ and another perpendicular to $\hat{\by}$), the structure of this broken-symmetry phase is an example of a non-chiral gyrotropic structure. This is unlike many of the examples of the OEE studied so far\cite{yoda2015current, yoda2018orbital}, such as trigonal selenium and tellurium\cite{rou2017kinetic, csahin2018pancharatnam}. Furthermore, it constitutes a concrete example of longitudinal magnetization induced by a current in a mirror-symmetric structure, while it has been implied that this is not allowed by symmetry~\cite{yoda2015current}. While longitudinal magnetization is forbidden by mirrors perpendicular or parallel to the current, it is not forbidden by mirrors at $45^\circ$ angles from the current.

\begin{figure}[tb]
\makebox[0.1\linewidth][c]{\includegraphics[width=0.67\linewidth]{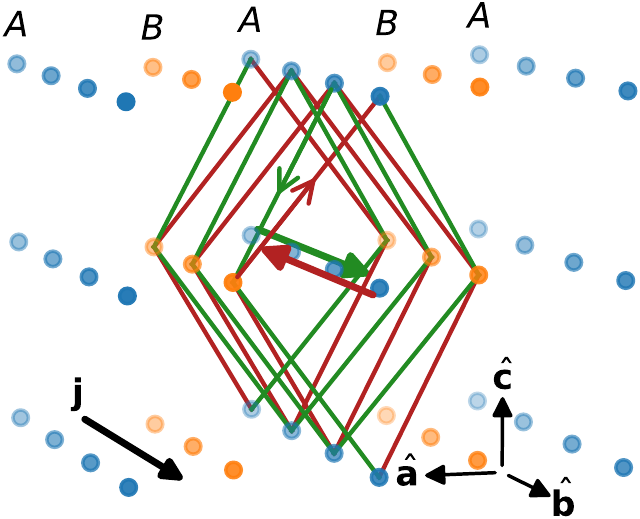}}
\caption{(Color online) For a current $\bj$ oriented in the $\hat{\bb}$ direction, solenoid-like paths due to the $t_{2A}$ hoppings (green) and the $t_{2B}$ hoppings (red) have opposite helicity. If $t_{2A}=t_{2B}$, the magnetizations from the two solenoids cancel. Crystal is not drawn to scale.}
\label{fig:solenoidsA}
\end{figure}

\begin{figure}[tb]
\makebox[0.1\linewidth][c]{\includegraphics[width=0.95\linewidth]{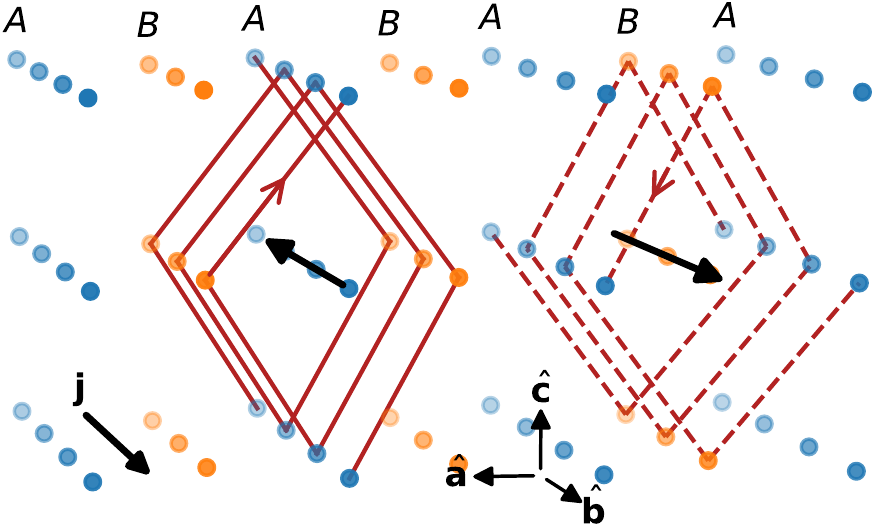}}
\caption{(Color online) For a current $\bj$ oriented in the $\hat{\bb}$ direction, solenoid-like paths with their axis along $A$ atoms (solid lines) have opposite helicity to those with their axis along $B$ atoms (dashed lines). If the atoms are indistinguishable, the magnetizations from the two solenoids cancel on average. Crystal is not drawn to scale.}
\label{fig:solenoidsB}
\end{figure}

Similarly to a previously studied model\cite{yoda2018orbital}, a picture of current flowing through solenoids provides a qualitative understanding of the OEE in phase I. Furthermore, this picture makes physically clearer why the OEE response vanishes (i) when $t_{2A}=t_{2B}$ --- hence clarifying physically the necessity of assuming $t_{2A}\neq t_{2B}$ from the start ---and (ii) in the absence of charge order, that is, when the atoms are indistinguishable. We imagine a current driven in the $\hat{\bb}$ direction: the simplest solenoid-like paths that result in net displacement purely in the $\hat{\bb}$ direction are shown in Fig.~\ref{fig:solenoidsA} and \ref{fig:solenoidsB}.

(i) Each row of atoms, whether on sublattice $A$ or $B$, is surrounded by a solenoid traced out of $t_{2A}$ (green) hoppings and another traced out of $t_{2B}$ (red) hoppings (Fig.~\ref{fig:solenoidsA}). These two solenoids are of opposite helicites, and if $t_{2A} = t_{2B}$, their induced magnetic fields cancel identically, precluding any current-induced magnetizationwhether in the symmetric phase or in a charge-ordered phase. Indeed, if $t_{2A}=t_{2B}$, there exist additional mirror planes (perpendicular to $\hat{\ba}$ and to $\hat{\bb}$) that interchange the two solenoids.

(ii) If $t_{2A}\neq t_{2B}$, let us presume without loss of generality that $t_{2B}$ (in red) is dominant and ignore the $t_{2A}$ solenoids. Figure~\ref{fig:solenoidsB} shows that a solenoid whose central axis is a row of $A$ sites is of opposite helicity as one whose central axis is a row of $B$ sites. If the atoms on the two sublattices are indistinguishable, the magnetic fields induced by these two solenoids are equal and opposite and cancel when averaged over several lattice spacings. If the $A$ and $B$ atoms are distinguishable, however, the two sets of solenoids are distinct and can give rise to net magnetization. Indeed, while in the symmetric phase, there exist glide planes perpendicular to $\hat{\ba}$ and along $\hat{\bb}$ that interchange the dotted-line and solid-line solenoids.

It is surprising \emph{a priori} that such a solenoid picture can exist in a nonchiral crystal (previous examples have focused on chiral structures\cite{yoda2015current, yoda2018orbital, csahin2018pancharatnam}); however, the two mirror symmetries of point group $\mathit{\bar{4}2m}$---one perpendicular to $\hat{\bx}$ and the other perpendicular to $\hat{\by}$---do not bring the solenoids into themselves, but rather exchange the $a$-axis solenoids with the $b$-axis solenoids. The mirrors instead explain why the OEE coefficients along the $a$ and $b$ axis are opposite in sign, since the current is a polar vector and the magnetization is an axial vector.

In summary, we have discussed a simple model of a crystal with lines nodes, and shown how charge order can lead to a nonzero OEE. Such an OEE could be probed by
nuclear magnetic resonance experiments. We have also discussed how the 
magnetoelectric response has a large contribution arising from the vicinity of line nodes. Our work suggests that lightly doped line-node semimetals in materials with weak 
spin-orbit coupling might be a promising place to search for large OEE.

\paragraph*{Note}
While completing this manuscript, we came across a recent theory preprint \cite{ishitobi2019magnetoelectric} which discusses the OEE induced by quadrupolar
symmetry breaking in certain diamond lattice materials.

\section*{Acknowledgements}
We thank D.~A.~Pesin for fruitful discussions. This research was funded by the Natural Sciences and Engineering Research Council of Canada and the Canadian Institute for Advanced Research. G.\ M. is supported by the \emph{Fonds de recherche du Qu\'ebec - Nature et technologies}. This research was enabled in part by support provided by Compute Ontario, Westgrid and Compute Canada. Computations were performed on the Niagara supercomputer at the SciNet HPC Consortium. SciNet is funded by: the Canada Foundation for Innovation; the Government of Ontario; Ontario Research Fund - Research Excellence; and the University of Toronto.

\appendix

\section{Classical phase diagram} \label{app:charge_orders}
\begin{figure}[tb]
    \centering
    \makebox[0.1\linewidth][c]{\includegraphics[width=1.12\linewidth]{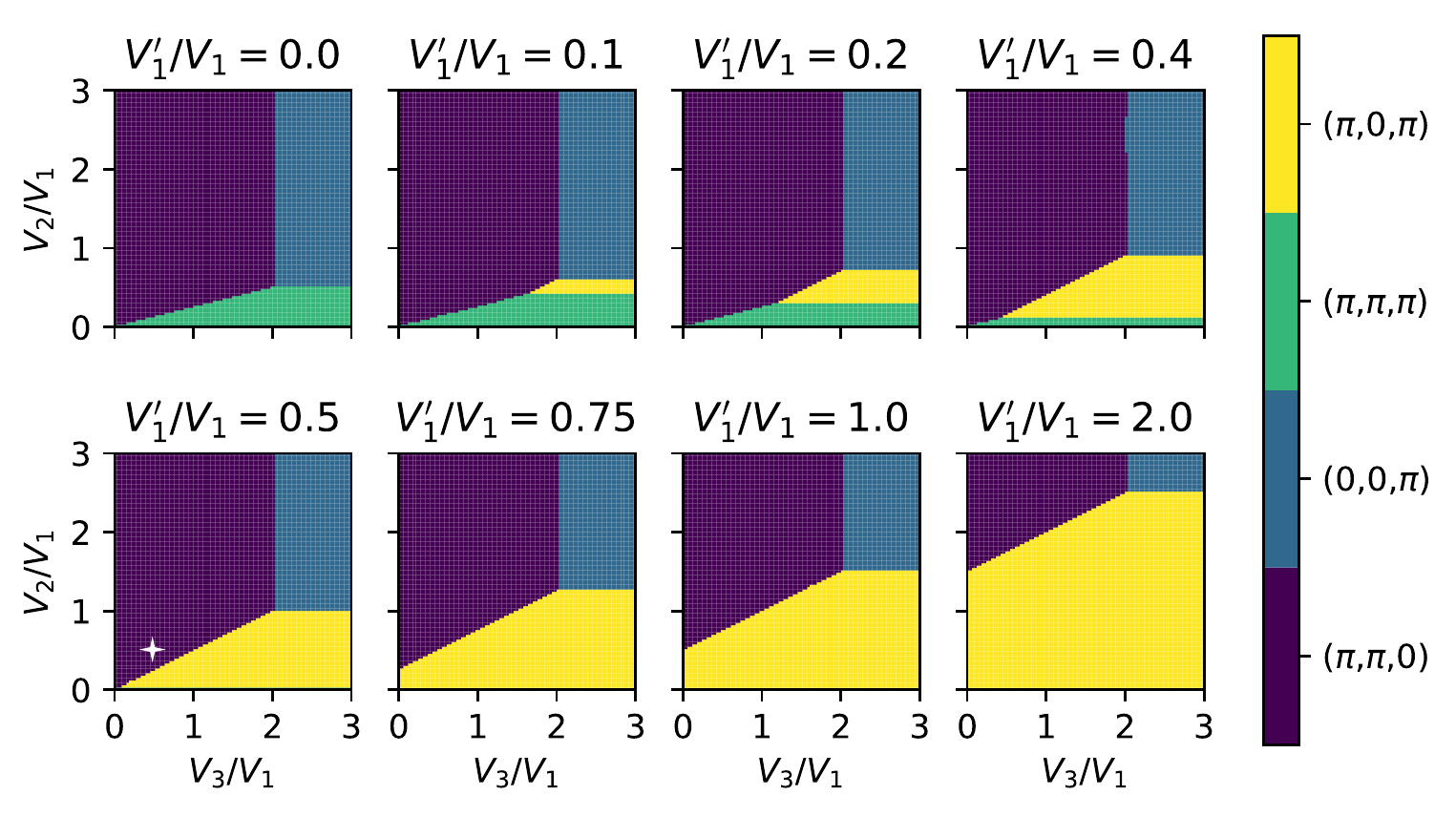}}
    \caption{(Color online) Phase diagram showing the lowest-energy mode (degenerate with symmetry-related modes) resulting from a model of electrostatic charges subject to the interaction of Eq.~\ref{eq:interaction}. Ordering wavevectors are expressed in $x$, $y$, and $z$ components with $a_0=c=1$. The white star shows the relative parameter values used in the MF calculation.}
    \label{fig:classical_phase_diagram}
\end{figure}

The ordering wavevectors favoured by the interaction of Eq.~\ref{eq:interaction} were identified by determining the energy of CDW modes according to $V$ in a classical picture of electrostatic charges. In Fig.~\ref{fig:classical_phase_diagram}, we show which mode has the lowest energy (according to $V$) as a function of the (relative) sizes of the repulsion strengths $V_1$, $V_1'$, $V_2$, and $V_3$. Note that, unlike in the main text, the modes are given in $x$-$y$-$z$ coordinates as $\big(Q_x, Q_y, Q_z\big)$ (where we have set $a_0=c=1$), since it is not necessary to work with a multi-atom basis when considering only the repulsion term $V$. Furthermore, note that the yellow phase $(\rpi,0,\rpi)$ implicitly stands for itself as well as its symmetry-related counterpart $(0,\rpi,\rpi)$; the two are degenerate as expected by symmetry.

The ansatz for the mean-field theory described in the main text was chosen to potentially allow all the ground states that occur in this simplified model.

\section{Details of the mean-field calculation} \label{app:MFT}
Here, we provide additional information regarding the self-consistent MF calculation of the CDW order. The interaction of Eq.~\ref{eq:interaction} was decomposed in the density channel, giving rise to the following MF interaction term:
\begin{subequations}
\begin{align}
V_\text{MF} &= \frac{1}{2} \sum_{(i,\alpha)\neq (j,\beta)} V_{ij}^{\alpha\beta} \left( \rho_i^\alpha n_j^{\,\beta} + n_i^\alpha \rho_j^{\,\beta} \right) \\
&= N \sum_{\bq}^\text{BZ} \sum_{\alpha\beta} 
%\re^{-\ri\bq\cdot(\bdelta^\alpha-\bdelta^\beta)} 
V_{\bq}^{\alpha\beta} {\rho_{\bq}^\alpha}^* n_{\bq}^{\,\beta}.
\end{align}
\end{subequations}
Here, $N$ is the number of unit cells in the crystal, $\rho_{\bq}^\alpha = N^{-1} \sum_{i} \re^{-\ri \bq \cdot \bR_i} \rho_i^\alpha$ (and likewise for $n_{\bq}^\alpha$), and $V_{\bq}^{\alpha\beta}$ is the Fourier transorm of $V_{ij}^{\alpha\beta}$, which is invariant under simultaneous translation of $i$ and $j$.

Choosing a closed set of commensurate wavevectors $\{\bQ\}$ defines a \emph{reduced Brillouin zone} (RBZ), which is mapped to the full Brillouin zone under addition of the wavevectors $\bQ$. This property allows us to rewrite the integral of reciprocal-space-periodic functions as
\begin{equation}
\sum_{\bk}^\text{BZ} f(\bk) = \sum_{\bQ} \sum_{\bk}^\text{RBZ} f(\bk + \bQ),
\end{equation}
where the domain of the momentum sums is indicated above the summation symbol. Hence, we diagonalized the Hamiltonian $H_\text{MF} = K + V_\text{MF}$ in the RBZ by writing it in the form
\begin{align}
K &= \sum_{\bk}^\text{RBZ} \sum_{\alpha\beta} \sum_{\bQ} 
{c^\alpha_{\bk+\bQ}}^\dagger h_{\bk+\bQ}^{\alpha\beta} c^{\,\beta}_{\bk+\bQ} \\
V_\text{MF} &= \sum_{\bk}^\text{RBZ} \sum_{\alpha} \sum_{\bQ \bQ'} 
{c^\alpha_{\bk+\bQ}}^\dagger \tilde{\rho}_{\bQ-\bQ'}^{\,\alpha} c^\alpha_{\bk+\bQ'},
\end{align}
where $\tilde{\rho}_{\bQ}^{\,\alpha} = \sum_{\beta} V_{\bQ}^{\alpha\beta} \rho_{\bQ}^{\,\beta}$ and $h_{\bk}^{\alpha\beta}$ is the Bloch Hamiltonian, in our case given in Eq.~\ref{eq:TB}: $h_{\bk} = d_{\bk}^0 + \vec{d}_{\bk}^{\vphantom{0}} \cdot \vtau$. Starting from a series of randomized values for the MFs $\rho_{\bQ}^\alpha$, we iterated until the computed expectation values agree with the input MFs to within $10^{-6}$. We compared the Helmholtz free energy $F$ of the different ground states thusly obtained and selected the one with minimal $F$ at every point in parameter space.

\end{document}